\documentclass[12pt,preprint]{aastex}
\usepackage{lscape}
\usepackage{amsmath}

\newcommand{\sn}{SN Ia }

\newcommand{\sne}{SNe Ia }
\newcommand{\snep}{SNe Ia}
\shorttitle{UV dispersion in HST nearby SNe Ia}
\shortauthors{Cooke et al.}

\begin{document} 

\title{Hubble Space Telescope Studies of Nearby Type Ia Supernovae:
The Mean Maximum Light Ultraviolet Spectrum and its Dispersion}
 
\author{Jeff Cooke\altaffilmark{1}, Richard S. Ellis\altaffilmark{1},
Mark Sullivan\altaffilmark{2}, Peter Nugent\altaffilmark{3}, D. Andrew
Howell\altaffilmark{4,}\altaffilmark{5}, Avishay
Gal-Yam\altaffilmark{6}, Chris Lidman\altaffilmark{7}, Joshua
S. Bloom\altaffilmark{8}, S. Bradley Cenko\altaffilmark{8}, Mansi
M. Kasliwal\altaffilmark{1}, Shrinivas R. Kulkarni\altaffilmark{1},
Nicholas M. Law\altaffilmark{9}, Eran O. Ofek\altaffilmark{1,10}, Robert
M. Quimby\altaffilmark{1} } \email{cooke@astro.caltech.edu.edu}

\altaffiltext{1}{Cahill Center for Astrophysics,California Institute of
Technology, Pasadena, CA 91125, USA}
\altaffiltext{2}{Department of Astrophysics, University of Oxford, UK}
\altaffiltext{3}{Computational Cosmology Center, Lawrence Berkeley 
National Laboratory, Berkeley, CA 94720, USA}
\altaffiltext{4}{Las Cumbres Observatory Global Telescope Network,
Goleta, CA 93117, USA}
\altaffiltext{5}{Department of Physics, University of California,
Santa Barbara, CA 93106-9530, USA}
\altaffiltext{6}{Astrophysics Group, Weizmann Institute of Science,
Rehovot 76100, Israel}
\altaffiltext{7}{Australian Astronomical Observatory, Epping, NSW
1710, Australia}
\altaffiltext{8}{Department of Astronomy, University of California,
Berkeley, CA 94720-3411, USA} 
\altaffiltext{9}{Dunlap Institute for Astronomy and Astrophysics,
University of Toronto, 50 St. George Street, Toronto M5S 3H4, Ontario,
Canada}
\altaffiltext{10}{Einstein Fellow}


\begin{abstract}
 
We present the first results of an ongoing campaign using the STIS
spectrograph on-board the Hubble Space Telescope (HST) whose primary
goal is the study of near ultraviolet (UV) spectra of local Type Ia
supernovae (SNe Ia).  Using events identified by the Palomar Transient
Factory and subsequently verified by ground-based spectroscopy, we
demonstrate the ability to locate and classify SNe Ia as early as 16
days prior to maximum light.  This enables us to trigger HST in a
non-disruptive mode to obtain near UV spectra within a few days of
maximum light for comparison with earlier equivalent ground-based
spectroscopic campaigns conducted at intermediate redshifts,
$\bar{z}\simeq0.5$.  We analyze the spectra of 12 Type Ia supernovae
located in the Hubble flow with $0.01<z<0.08$.  Although a fraction of
our eventual sample, these data, together with archival data, already
provide a substantial advance over that previously available.
Restricting samples to those of similar phase and stretch, the mean UV
spectrum agrees reasonably closely with that at intermediate redshift,
although some differences are found in the metallic absorption
features.  A larger sample will determine whether these differences
reflect possible biases or are a genuine evolutionary effect.
Significantly, the wavelength-dependent dispersion, which is larger in
the UV, follows similar trends to that observed at intermediate
redshift and is driven, in part, by differences in the various
metallic features.  While the origin of the UV dispersion remains
uncertain, our comparison suggests that it may reflect compositional
variations amongst our sample rather than being predominantly an
evolutionary effect.  \end{abstract}



\section{INTRODUCTION}\label{intro}

Type Ia supernovae (\snep) remain the most practical and
well-exploited cosmological probe offering an immediate route to
understanding `dark energy'.  Measures of distant events are being
used to distinguish between Einstein's cosmological constant,
$\Lambda$, and a scalar field whose equation of state parameter
$w\neq$$-1$ \citep{astier06,riess07,kessler09,amanullah10}.  Yet
despite remarkable observational progress, there is no satisfactory
theory explaining a \sn event.  The mechanism by which a white dwarf
accretes additional material is unclear as is the nature of the
explosion itself \citep{livio99}.

To facilitate progress, observers employ a variety of empirical
correlations to reduce the intrinsic scatter of the \sn Hubble
diagram.  \sne were initially considered a one or two parameter family
with the light curve width and rest-frame color as the key variables.
However improved data have revealed important correlations with the
host galaxy.  Events are not only more common in star-forming hosts
per unit stellar mass but their light curve properties differ from
those seen in quiescent galaxies \citep{sullivan06b,sullivan10}, an
effect that has direct consequences for their use over large look-back
times \citep{howell07}.

These discoveries naturally raise the question of what further
evolutionary changes might be present in the \sn population.  A
long-standing concern has been the unknown effect of an {\it evolving
progenitor composition}, both in terms of a possible
redshift-dependent bias, and in producing an intrinsic dispersion that
could limit the effectiveness of large future surveys.  A Keck study
of high quality rest-frame near-UV \sne spectra at intermediate
redshift ($z\simeq0.5$) reveals a surprising diversity at short
wavelengths where some models predict a sensitivity to metallicity
\citep[][hereafter E08, see also Foley et al. 2008]{ellis08}.  A large
U-band dispersion had earlier been claimed in the local photometric
survey of \citet{jha06}, although subsequent photometry has challenged
the amount \citep{astier06}.  Although some models predict a UV
dispersion might arise from variations in the progenitor composition
\citep{lentz00,hoef00}, the magnitude of the effect seen by E08
exceeds that predicted for reasonable compositional differences.
\citet{sauer08} have shown a large fraction of the UV flux can be
formed by reverse-fluorescence scattering which affects the dependence
on composition. If the observed dispersion is found to arise from some
evolutionary trend, it could bias future $z>1$ SNe campaigns that
typically sample from this wavelength region.  \citet[][hereafter
S09]{sullivan09} compare mean \sn spectra over a redshift path of
$0<z<1.2$ and find no strong evolution, however, only three local UV
spectra were available at the time, seriously limiting the comparison
and giving no local measure of the UV dispersion.

Efforts to understand the UV behavior of \sne have subsequently
intensified.  Using the UV optical telescope on-board SWIFT,
\citet{brown10} and \citet{milne10} have confirmed the presence of a
dispersion increase to shorter wavelengths in local SNe Ia.
Comparisons between local and intermediate redshift data remain
confused and so the question remains as to whether some component of
the significant diversity of UV spectra seen in distant \sne is an
evolutionary phenomenon or represents some as yet unexplained
diversity in the \sn mechanism.

As described in E08, spectroscopic studies offer a major advantage
over photometric investigations as they eliminate uncertainties
arising from k-corrections and, with adequate data, the mean and
dispersion can be investigated in the context of known metallic
features.  Following the successful repair of the UV-capable Space
Telescope Imaging Spectrograph (STIS) aboard the Hubble Space
Telescope ($HST$) during the 2009 Servicing Mission 4, it has become
possible to make significant progress in addressing the above
questions.  Using a non-disruptive Target of Opportunity campaign (GO
11721, PI: Ellis), we are securing maximum light STIS spectra for 35
\sne located in the Hubble flow.  Here we present the results from the
first 12 events from this program, augmented by three events from
earlier archival data, that provide comparable statistics to the
survey of E08 at intermediate redshift for the analysis undertaken
here.


\begin{deluxetable}{lccl}
\tabletypesize{\normalsize}
\tablecaption{Low Redshift Supernova Ia Sample
\label{details}}
\tablewidth{0pt}
\tablehead{\colhead{Supernova} & \colhead{Phase\tablenotemark{a}} &
\colhead{Stretch} & \colhead{Host z}} 
\startdata
PTF-09dlc & $+2.02\pm0.38$ & $1.152\pm0.054$ & 0.0675\\
PTF-09dnl & $+0.68\pm1.99$ & \ldots        & 0.0231\\
PTF-09dnp & $+5.47\pm0.87$ & $0.996\pm0.308$ & 0.0373\tablenotemark{c,d}\\
PTF-09fox & $+1.66\pm0.40$ & $1.016\pm0.109$ & 0.0718\\
PTF-09foz & $+2.59\pm0.38$ & $0.883\pm0.079$ & 0.0543\tablenotemark{c,d}\\
PTF-10bjs & $+2.10\pm0.18$ & $1.138\pm0.019$ & 0.0300\tablenotemark{c,d,e}\\
PTF-10fps & $+8.15\pm4.85$ & $0.980\pm0.530$ & 0.0215\tablenotemark{c,d}\\
PTF-10hdv & $+3.10\pm0.36$ & $1.077\pm0.064$ & 0.0533\\
PTF-10hmv & $+2.73\pm0.09$ & $1.150\pm0.009$ & 0.0324\\
PTF-10icb & $+0.76\pm0.13$ & $1.071\pm0.021$ & 0.0086\tablenotemark{c,d}\\
PTF-10mwb & $-0.19\pm0.14$ & $0.896\pm0.018$ & 0.0313\\
SN 2009le & $-0.32\pm1.99$ & $\dots$ & 0.0178\tablenotemark{d}\\
\hline
SN 1981b  & $+1.30\pm0.14$ & $0.89\pm0.02$ & 0.0060\tablenotemark{f}\\
SN 1992a  & $+5.45\pm0.04$ & $0.82\pm0.01$ & 0.0061\tablenotemark{f}\\
SN 2001ba & $+4.21\pm0.22$ & $1.02\pm0.02$ & 0.0305\tablenotemark{f}
\enddata
\tabletypesize{\scriptsize}
\tablenotetext{a}{Effective phase (phase/stretch) at the time of the
  STIS observations }
\tablenotetext{b}{Unless otherwise noted, redshifts are obtained from
  the host features in the PTF spectra and are accurate to $z\lesssim0.001$}
\tablenotetext{c}{Sloan Digital Sky Survey (SDSS)}
\tablenotetext{d}{NASA Extragalactic Database (NED)}
\tablenotetext{e}{May reside in a possible line-of-sight dwarf galaxy at
  $z=0.019$}
\tablenotetext{f}{Archival UV spectra discussed by S09
  \citep{branch83,kirshner93,foley08}}
\end{deluxetable}
\normalsize

\section{OBSERVATIONS}\label{data}

\begin{figure}
\begin{center}
\scalebox{0.75}[0.75]{\rotatebox{90}{\includegraphics{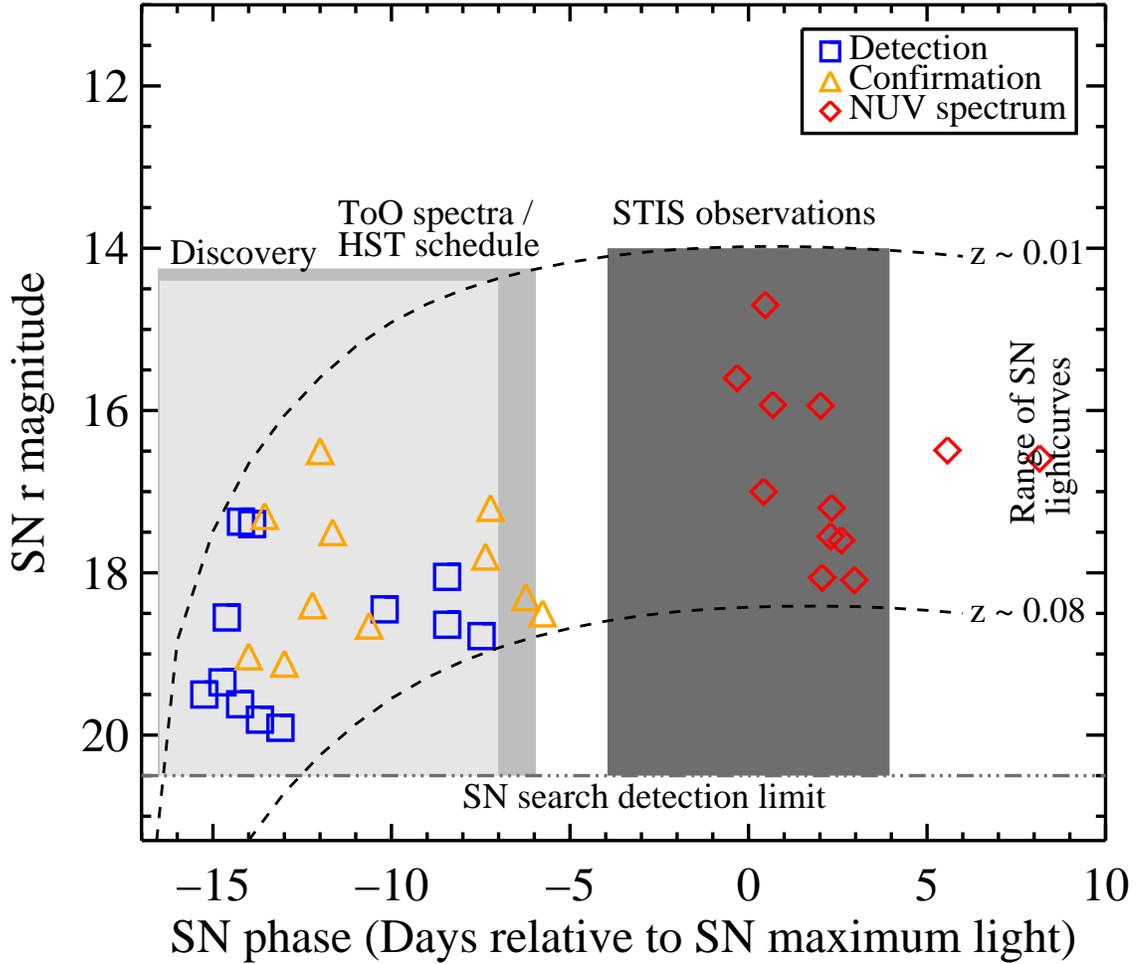}}}
\caption
{\small Timeline for the HST non-disruptive ToO program (GO 11721, PI:
  Ellis).  The magnitude and phase (number of days relative to maximum
  light) are shown for the PTF, and one non-PTF, photometric 
  discoveries (squares), ground-based spectroscopic confirmations 
  (triangles), and near-UV STIS spectra (diamonds).  Photometric 
  discoveries and spectroscopic confirmation are necessary during the
  time windows indicated by the light-gray and gray block regions, 
  respectively, in order for STIS near UV spectra to be acquired within 
  the time window indicated by the dark-gray region. }
\label{program}
\end{center}
\end{figure}

A significant challenge in delivering targets to HST for observations
at maximum light is the need to detect and identify convincing SNe Ia
candidates soon after explosion.  For a {\it non-disruptive} Target of
Opportunity (ToO) campaign, defined as one where Phase II observations
are submitted for inclusion in the HST schedule built for the second
week following the submission, SN confirmation must occur $\sim10-12$
calendar days prior to submission.  As a result, spectroscopic SNe Ia
target confirmations $\sim7-16$ days before maximum light are
necessary to acquire STIS observations with phases $\pm4$ days from
maximum light and match a distribution of intermediate-redshift 
near UV spectra secured by the E08 campaign.
Figure~\ref{program} illustrates the timeline of the data acquisition
for the 12 SNe of our HST program presented here -- from detection, to
ground-based spectroscopic confirmation, to STIS observation -- and
shows the ability of PTF to detect SN Ia outbursts as early as $16$
days prior to maximum light.

The identification of SNe Ia suitable for HST non-disruptive
observations proceeded in two stages.  The early identification of
candidate events was based on photometric survey data.  Eleven of the
12 newly-discovered SNe were identified between 2009 August 19 and
2010 June 29 by the Palomar Transient Factory
\citep[PTF;][]{rau09,law09}.  As with the CFHT Supernova Legacy Survey
\citep[SNLS][]{astier06} utilized by the E08 intermediate redshift
program, PTF is a rolling search for transient events unbiased to the
nature of the host galaxy.  The remaining event (SN 2009le) was
triggered on the reported results of an independent search
\citep[][CHASE]{pignata09} during a period when the Palomar
observatory was under extensive ash clouds from the 2009 California
forest fires.

\begin{figure}
\begin{center}
\scalebox{0.90}[0.90]{\includegraphics{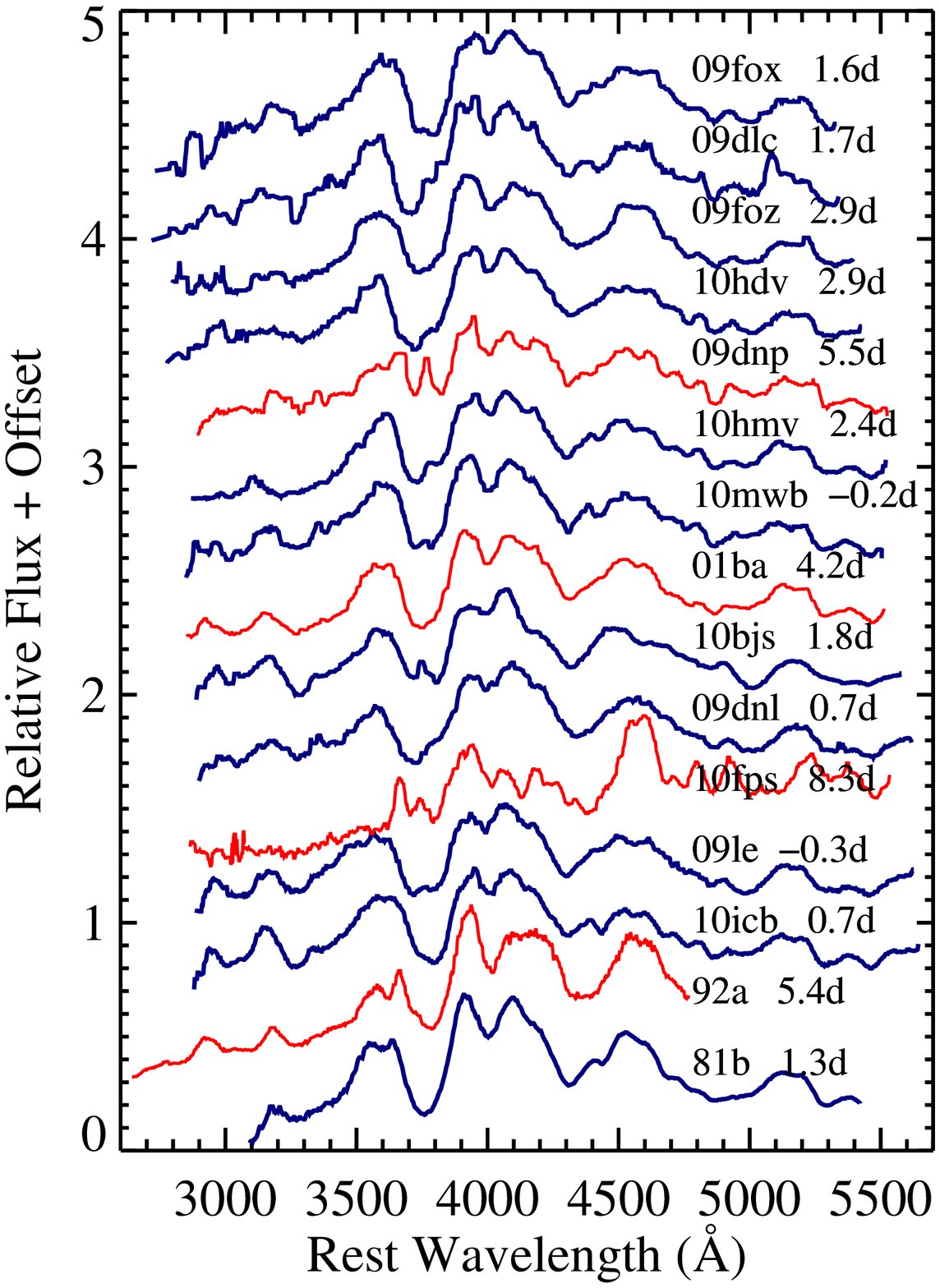}}
\caption
{\small STIS near UV maximum-light spectra as itemized in Table
  \ref{details} reduced to rest-frame wavelengths.  Each label gives
  the PTF identification (with 09le indicating SN 2009le) and phase 
  of STIS observations.  Spectra in red represent those not included 
  in the mean and dispersion analysis (see text).}
\label{stis_spec}
\end{center}
\end{figure}

Ground-based spectroscopic follow-up acquired within one to a few days
after photometric detection represents the second stage of
confirmation essential for determining the SN type and redshift.
Spectroscopy was performed using regularly scheduled time and
occasional ToO interrupts on the following telescopes/instruments: the
Low-Resolution Imaging Spectrometer \citep[LRIS;][]{oke95,mccarthy98}
and DEep Imaging Multi-Object Spectrograph \citep[DEIMOS;][]{faber03}
on the Keck telescopes, the Gemini Multi-Object Spectrograph
\citep[GMOS;][]{hook04} on the Gemini telescopes, FORS
\citep{appenzeller98} and X-shooter \citep{vernet09} on the ESO Very
Large Telescopes, the Double Spectrograph \citep[DBSP;][]{oke82} on
the Palomar Hale Telescope, and the Intermediate dispersion
Spectrograph and Imaging System (ISIS) on the William Herschel
Telescope.  By matching flux-calibrated spectra with selections from
an extensive spectral database \citep{howell05}, improved phases
typically accurate to $\pm2$ days were obtained.

Near-UV spectra were acquired using the Space Telescope Imaging
Spectrograph \citep[STIS;][]{woodgate98,kimble98} between 2009
September 02 and 2010 July 13 (Fig.~\ref{program}, red diamonds).  To
best match our $z\simeq0.5$ Keck spectra (E08), we used the STIS/CCD
430L observing mode with $3-4$ CR-SPLIT exposures totaling 1 orbit,
ensuring good signal-to-noise ratio coverage down to a rest wavelength
below $2900$\AA.  A montage of the STIS spectra is shown in
Figure~\ref{stis_spec}.  Associated near-simultaneous ground-based
optical spectra were also taken using further regularly scheduled time
and ToO interrupts using the facilities described above.  All data
were reduced using standard IRAF and IDL data reduction routines.
Pre- and post-maximum light PTF $r$-band photometric observations were
used to determine the initial phase and stretch for each $HST$
triggered event using the $SiFTO$ light curve fitter \citep{conley08}.
Details for the SN Ia sample here and the three archival SNe Ia
(discussed in S09) are listed in Table~\ref{details}.

In 4 cases (09dnl, 09dnp, 09fox, 10fps), the PTF light curves are of
marginal quality or too poorly sampled for an accurate determination
of one or both of the phase and stretch.  One event (SN 2009le) has no
PTF photometry and its phase was determined from the CHASE estimate
\citep{challis09} which is consistent with our spectroscopic estimate.
In our final analysis (\S3), this imprecision affects the discussion
of two of our STIS spectra (09dnl, 2009le).

\section{ANALYSIS}\label{analysis}

\begin{figure}
\scalebox{0.75}[0.75]{\rotatebox{90}{\includegraphics{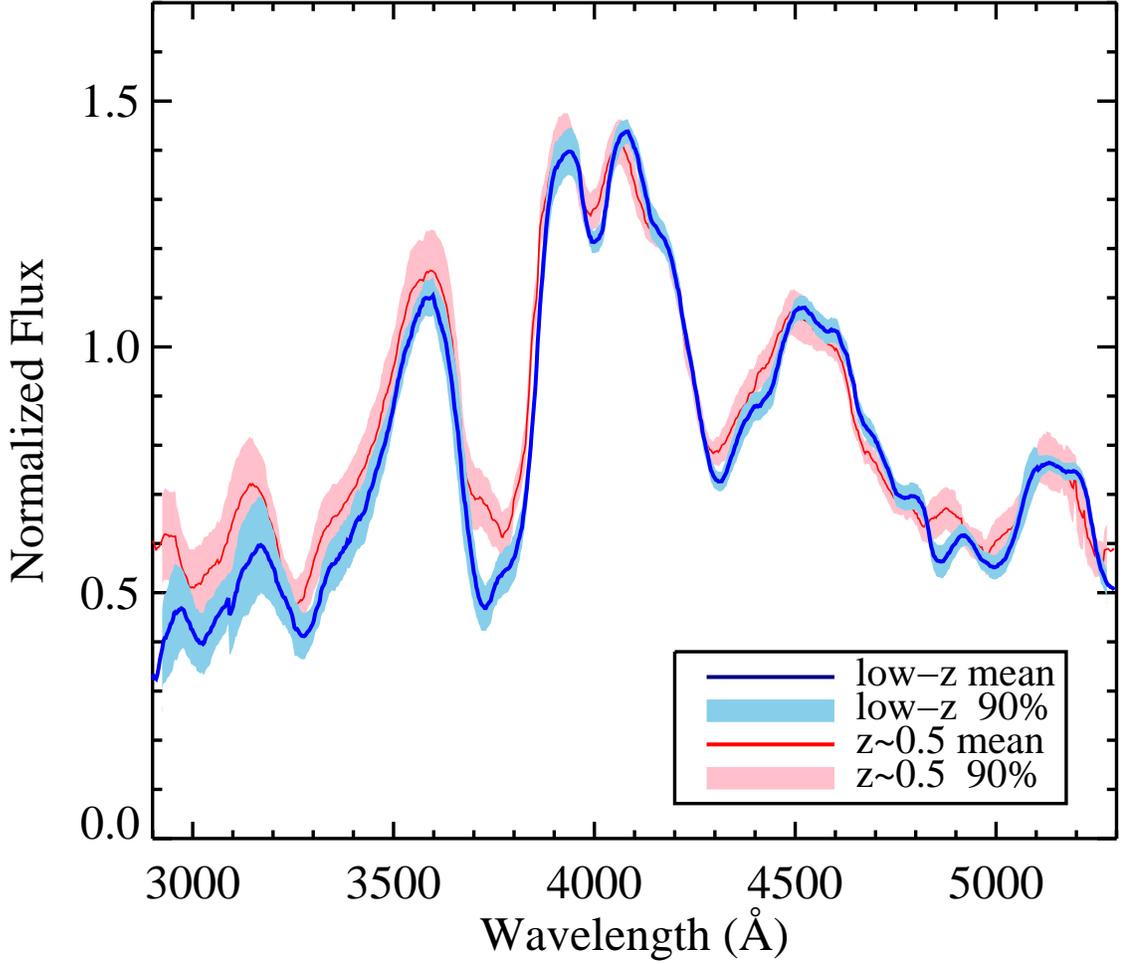}}}
\caption
{\small Mean near-UV spectra for low- and intermediate-redshift SN 
  Ia near maximum light.  The mean near-UV spectrum derived from the
  10 STIS supernovae and one archival SN compliant with the adopted 
  phase and stretch criteria (see text) is shown by the blue line 
  with the region containing 90\% of the jack-knife resampling fits 
  shown as the light blue region.  This is compared to the mean of a 
  $z\simeq0.5$ E08 sample of $16$ SNe closely matched in phase and 
  stretch (red line) and its 90\% region (bounded in pink). } 
\label{best} 
\end{figure}

Prior to HST Cycle 13, UV spectra were secured for 13 local SNe\,Ia,
but only 5 were studied close to maximum light and 2 of these were
peculiar.  Those suitable from earlier work are listed in
Table~\ref{details} and shown in Figure~\ref{stis_spec}.  In Cycle 13
a ToO campaign began to increase the sample (GO 10182, PI: Filippenko)
but the failure of STIS curtailed this program.  Remarkably, prior to
our study, more was known about the pre-maximum and maximum light UV
spectra of SNe\,Ia at $z\sim0.5$ than at $z=0$.  Our STIS sample has
now dramatically improved this situation.

In order to compare the mean UV spectra of local SNe\,Ia and their
dispersion with the sample discussed by E08, we must match the phase
(and ideally the stretch) distributions of the two samples.  Such a
matched comparison was not possible in S09 due to the paucity of the
local data.  To maximize the utility of the new HST data, we adopt a
phase range of $-0.32$ to $+4$d (Figure~\ref{program}) bounded by the
earliest phase of the HST data ($-0.32$) and a $+4$d phase limit
similar to that applied in E08 to minimize phase evolution effects on
the mean and dispersion.  This criterion leads to 11 events from the
HST+archival sample drawn from Table~\ref{details} and 16 events from
E08.  Ignoring the two events for which accurate stretches cannot be
determined, the mean values are $\bar{S_{HST}}= 1.030\pm0.038$ and
$\bar{S_{z\sim0.5}}= 1.049\pm0.019$ which are consistent within the
errors.  Excluding the two supernovae for which reliable stretches are
not available does not significantly affect our results in the
following subsections.

\subsection{Mean Type Ia Spectrum}\label{test}

S09 presented a comparison of the mean UV spectra determined by E08 at
$z\simeq0.5$, those secured using the ACS grism by \citet{riess04} at
$z\simeq1.2$, and three local spectra from archival data listed in
Table~\ref{details}.  Examining the $z\simeq0.5$ UV spectra, S09 found
some decrease with redshift in the strength of intermediate mass
element features (Si \textsc{ii}, Ca \textsc{ii}, and Mg \textsc{ii}),
but it was argued this could arise in part due to the natural drift to
luminous, larger stretch events expected at high redshift. Below
$\lambda\simeq$ 3600 \AA\ the mean local spectrum was highly
uncertain.

To facilitate a proper comparison to S09, we construct our mean
near-UV spectrum following the procedure discussed in E08.  Briefly,
the spectra are normalized to have the same flux through a box filter
defined between rest-frame $4000-4800$\AA~and the variation in the
mean spectrum is estimated via bootstrap-resampling.  Use of other box
filters, including the full wavelength range common to all spectra,
does not significantly affect the results.  Figure~\ref{best} shows
the normalized mean spectrum and region containing $90$\% of 100
bootstrap-resampled mean spectra for both the low- and
intermediate-redshift matched samples.  Although the low redshift SN
Ia mean spectrum closely resembles that at $z\simeq0.5$, as in S09, we
notice a marked decrease with increasing redshift in the depth of the
Si \textsc{ii} and Ca \textsc{ii} blend near 3800 \AA, Mg \textsc{ii}
near 4300 \AA, and those of iron group elements below $3500$\AA.  The
utility of the comparison is more advantageous than that conducted by
S09 for wavelengths below 3500\AA\ since our new local sample is much
larger than that used by S09 and the phase and stretch distributions
are better matched.  Farther into the UV we note that the differences
between the mean local and $z\simeq$0.5 spectra become particularly
significant.  Since the stretch distributions of the two samples are
similar, this could be a genuine effect rather than one arising from
samples biased to more luminous and bluer events \citep{balland09}.
Analysis of our eventual full sample will clarify this important
point.

\begin{figure}
\scalebox{0.75}[0.75]{\rotatebox{90}{\includegraphics{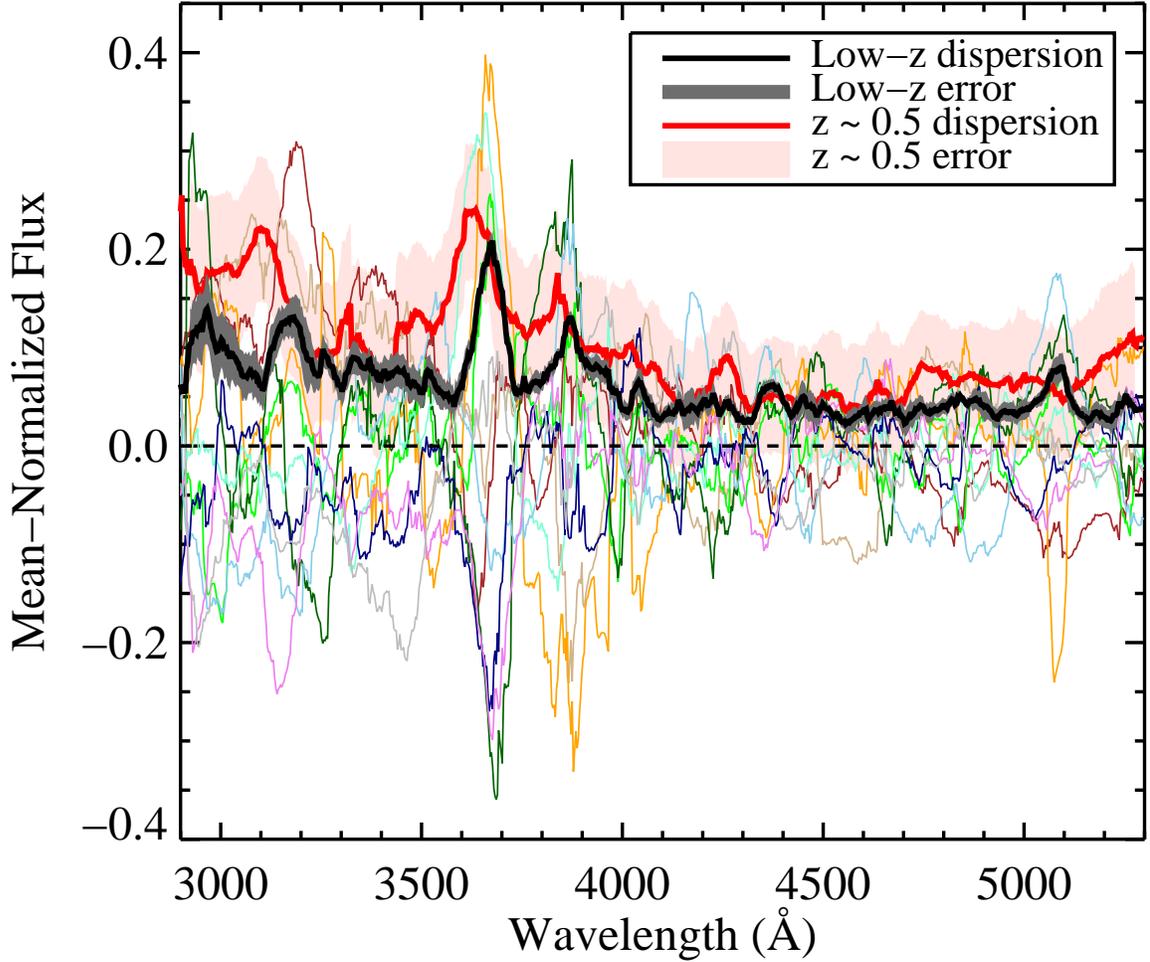}}}
\caption
{\small Dispersion from the mean for the 11 low-redshift SNe meeting 
  our phase criteria (multi-colored thin curves).  The absolute 
  value of the mean dispersion is indicated by the thick black curve 
  with $1\sigma$ observational uncertainties overlaid (gray region).  
  For direct comparison, the absolute value of the mean dispersion for 
  the matched sample of $16$ intermediate redshift E08 SNe is shown 
  (thick red curve).  An increased dispersion at shorter wavelengths
  is present in both samples. }
\label{disp}
\end{figure}

\subsection{Spectral Dispersion}\label{criteria}

E08 demonstrated a significant increase in the variance of their
$z\simeq$0.5 SNe Ia UV spectra for wavelengths below $3700$\AA, both
by comparing individual deviations of $15$ maximum light spectra from
their mean in units of the dispersion $\sigma$, and via photometric
colors measured directly from color-corrected Keck spectra.  For the
first time, our low-redshift sample of UV SN Ia spectra is large
enough to perform a similar analysis.  Ideally one would color-
correct the local spectra following the procedure discussed in E08
using the SALT2 color law \citep{guy07} but this requires host-
corrected multi-color data that must await late-time reference images
in bands other than the PTF $r$-band.  We experimented with estimating
the host contamination in $g$ and $i$ from the contemporary data but
concluded the uncertainties are too great at this time.  Accordingly,
a comparison with data from E08 that is not color-corrected is more
appropriate.  It is important to note that the color correction E08
applied did not significantly change the UV scatter and its wavelength
dependent trend.  The key question we seek to address is whether the
dispersion trend is generic to all SNe Ia, independent of redshift, or
largely a feature {\it of the intermediate redshift data only},
possibly implying some evolutionary effect.

Figure~\ref{disp} shows that the wavelength-dependent dispersion is
indeed present in the local data, consistent with the photometric
claims of \citet{brown10} and \citet{milne10}.  In the region that
contains 90\% of the mean values from bootstrap resampling, the
variation from one spectrum to another with respect to the mean
increases below $3700$\AA\ as in E08.  Clearly several of the features
which vary between the local and E08 samples discussed earlier
contribute to the dispersion suggesting a compositional origin.

Even if we exclude the region dominated by strong features and
consider the average dispersion from the mean spectrum in the regions
UV: 2900-3500 \AA\ and optical: 4100-5200 \AA\ allowing for the
spectrophotometric uncertainties, we find that the dispersion
increases from the optical to the UV by a similar factor of
$\simeq$2.3 in both the local and $z\simeq$0.5 samples.  There is
marginal evidence that the UV scatter may be larger in the
$z\simeq$0.5 data than in the local sample, but the spectrophotometric
uncertainties for the individual spectra are naturally larger.
Overall, we conclude that the UV spectral dispersion is most likely a
feature of the SNe Ia population and not an evolutionary effect.


\section{DISCUSSION}\label{disc}

Our initial results clarify and quantify indications from earlier
work.  Although the SN Ia mean spectrum close to maximum light appears
to have remained remarkably similar over the past 5 Gyr, we find the
decrease in the strength of the metallic features with increasing
redshift noted by S09 also present in our more representative
comparison.  Given that the mean stretches of the local and
z$\simeq$0.5 samples are similar, this may represent the expected
decrease in metallicity over this epoch.  We can address this
possibility in more detail with the completion of our survey.

Equally important is that we observe a strong wavelength dependent
scatter in the rest-frame UV spectra of our local sample, as noted in
E08.  Independent of the calibration questions that have plagued
recent photometric studies, the spectra demonstrate that the UV
scatter is generic to SNe Ia over a wide range of cosmic time and is
not likely an evolutionary effect.  Much of this behavior can be
attributed to the varying absorption line strengths of intermediate
mass elements occupying the UV wavelength region, supporting the
notion that the UV scatter arises from compositional differences
between events.

In addition to strengthening these conclusions with a larger sample,
it is now clear that further progress will follow more detailed
multi-phase UV studies of selected local events.  HST has recently
been awarded for such a program (GO 12298, PI: Ellis).

\acknowledgments

RSE acknowledges support from DOE grant DE-SC0001101, MS from the
Royal Society, A.G. from the Israeli Science Foundation and a European
Union Marie Curie fellowship.  Support for program GO 11721 was
provided by NASA through a grant from the Space Telescope Science
Institute, which is operated by AURA Inc. under NASA Contract
NAS5-26555.

\end{document}